\documentstyle[aps,multicol,prl,epsfig] {revtex}
\draft
\begin{document}
\def \bm #1 {\mbox{\boldmath $#1$}}
\title { Estimation of Initial Conditions and Secure Communication}
\author{ Anil Maybhate\footnote{e-mail: nil@prl.ernet.in}$^{1,2}$, 
         R.E.Amritkar\footnote{e-mail: amritkar@prl.ernet.in}$^1$ and
         D. R. Kulkarni\footnote{e-mail: drkul@prl.ernet.in}$^1$.}
\address{$^1$Physical Research Laboratory,Navrangpura,Ahmedabad 380009, India}
\address{$^2$Department of Physics,University of Pune,Pune 411007, India}
\maketitle

\begin{abstract} 
We estimate the initial conditions of a multivariable
dynamical system from a scalar signal, using a
modified Newton-Raphson method incorporating the time evolution.
We can estimate initial conditions of periodic and
chaotic systems and the required length of scalar signal is very small.
We also find that the information flow from one variable to
the other has logarithmic dependence on time. An important application
of the method is in secure communications. The communication procedure 
has several advantages as compared to others using dynamical systems.
\end{abstract}

\begin{multicols}{2}

A trajectory of a given dynamical system in its state space
depends on the set of initial conditions with which it starts.
In particular, the state of a chaotic system at a later time is
exponentially sensitive to changes in its initial state~\cite{Dev}. This
feature of a chaotic system leads to a complex behaviour in state
space that appears random yet is deterministic, and the time evolution 
is uniquely fixed by the initial state of the system. Though
there are several invariant measures of a chaotic system which are not
sensitive to the initial conditions, the exact trajectory crucially depends
on the initial state and hence is difficult to reproduce due to sensitivity
to initial conditions.

In light of these facts, it is interesting and important to ask whether
the complete set of initial conditions of a given multivariable dynamical
system can be estimated from a given scalar time series of a single state
space variable. In this letter we present a novel and simple
method to estimate the initial conditions from a given scalar time
series. The method is based on a modified multidimensional
Newton-Raphson method~\cite{Dev,PTVF} that includes the time evolution
of the system.  The length of the time series required for estimating the
initial conditions is very small. Also, the method works even when the 
conditional Lyapunov exponents are positive.

An important application of our method is in the area of secure
communications~\cite{CO,HGO}. A nice feature of this application is that the signal
that is transmitted is not the one modulated by the information
signal, making it difficult to crack the method. Our method is also
useful in the problem of synchronization of chaotic signals~\cite{PC1}. The
knowledge of initial conditions means the response system can be be
synchronized with the drive system almost instantaneously thereby
removing the problem of transients. Also synchronization can be
achieved in most of the cases where other methods fail~\cite{PC1}.

Let us consider an autonomous dynamical system given by,
\begin{equation}
{\bf\dot x} = {\bf F}({\bf x}),
\label{SYS}
\end{equation}
where ${\bf x}=(x_1,x_2,\dots,x_d)$ is a $d$-dimensional state vector
whose evolution is governed by the function ${\bf F} = (F_1, F_2, \dots,
F_d)$. Given an initial state vector ${\bf x}(0)$ at time $t=0$, the time
evolution ${\bf x}(t)$ is uniquely determined by Eq.~(\ref{SYS}). Now let
us assume that only one component of the state vector is known to us and
we take it to be $x_1(t)$ without loss of generality. The problem that we
address is to obtain the initial state vector ${\bf x}(0)$ from the
knowledge of the scalar signal $x_1(t)$.

Let ${\bf y}(0)$ denote a random initial state vector and ${\bf y}(t)$ its
time evolution obtained from Eq.~(\ref{SYS}). Let ${\bf w}(t)$ denote the
difference ${\bf w}(t) = {\bf y}(t) - {\bf x}(t)$.
We look for the solution of the equation
\begin{equation}
{\bf w}(t) = 0.
\label{wZERO}
\end{equation}
Noting that the initial state vectors ${\bf y}(0)$ and ${\bf x}(0)$
uniquely determine the difference ${\bf w}(t)$, one of the solutions of
Eq.~(\ref{wZERO}) is ${\bf y(0)} - {\bf x}(0)=0$ and this is the solution
that we are searching for.

We now introduce the notation ${\bf w}^n={\bf w}^n({\bf y}^0,{\bf
x}^0)={\bf w}(n\Delta t)$, where $\Delta t$ is a small time interval.
Similarly, ${\bf y}^n = {\bf y}(n\Delta t)$ and ${\bf x}^n = {\bf
x}(n\Delta t)$. With this notation condition~(\ref{wZERO}) can be written
as ${\bf w}^n = 0$.

Our approach to the solution of Eq.~(\ref{wZERO}) is a modified
Newton-Raphson method~\cite{Dev} which includes the time evolution of the
system.

Let us first consider ${\bf w}^1$. We have
\begin{eqnarray}
0 &=& {\bf w}^1({\bf x}^0, {\bf x}^0) =
            {\bf w}^1({\bf y}^0 + \delta {\bf y}^0, {\bf x}^0), \nonumber \\
&=&  {\bf w}^1({\bf y}^0, {\bf x}^0) 
+(\delta {\bf y}^0 \bm{\cdot \nabla}_{{\bf y}^0} )
{\bf w}^1({\bf y}^0,{\bf x}^0) + {\cal O}((\delta {\bf y}^0)^2),
\label{taylorx}
\end{eqnarray}
where $\delta {\bf y}^0 = {\bf x}^0 - {\bf y}^0 = - {\bf w}^0$ and the
last step is a Taylor series expansion in $\delta {\bf y}^0$. For small
$\Delta t$, we can write
\begin{equation}
{\bf w}^1({\bf y}^0, {\bf x}^0)
= {\bf w}^0 + \Delta t \, [{\bf F}({\bf y}^0)-{\bf F}({\bf x}^0)]
+ {\cal O}((\Delta t)^2).
\label{taylort}
\end{equation}
Substituting Eq.~(\ref{taylort}) in Eq.~(\ref{taylorx}) and neglecting
higher order terms, we get,
\begin{equation}
{\bf w}^1({\bf y}^0, {\bf x}^0) = {\bf w}^0 - \Delta t ({\bf w}^0
\bm{\cdot \nabla}_{{\bf y}^0} ) {\bf F}({\bf y}^0).
\label{veceq}
\end{equation}
It is convenient to write the above equation in a matrix form as
\begin{eqnarray}
W^1 &=& (I + \Delta t \, J^0) W^0 = A^0 W^0,
\label{W1}
\end{eqnarray}
where $W^n$ is the column matrix corresponding to the vector ${\bf w}^n$,
$I$ is the identity matrix, $A^n = I + \Delta t \, J^n$, and the elements
of the Jacobian matrix $J^n$ are
$J^n_{ij} = \partial F_i({\bf y}^n) / \partial y_j^n$.

Next we consider ${\bf w}^2$ or $W^2$. Proceeding as above, we get (see
Eq.~(\ref{W1})),
\begin{eqnarray}
W^2 & = & (I + \Delta t \, J^1) W^1 = (I + \Delta t \, J^1)(I + \Delta t \, J^0) W^0 \nonumber \\
& = & A^1 A^0 W^0.
\label{W2}
\end{eqnarray}
Similarly, the equation for $W^n$ is
\begin{eqnarray}
W^n & = & A^{n-1} A^{n-2} \cdots A^0 W^0.
\label{Wn}
\end{eqnarray}

We now concentrate on the first component of the signal whose time series
is assumed to be known. For a $d$-dimensional system we need $d-1$
equations to determine the initial state vector ${\bf x}^0$.
Eqs.~(\ref{W1}), (\ref{W2}) and~(\ref{Wn}) give us the required relations.
\begin{eqnarray}
W^1_1 & = & \sum_{i=1}^d A^0_{1i} W^0_i, \nonumber \\
W^2_1 & = & \sum_{i,j=1}^d A^1_{1i} A^0_{ij} W^0_j, \nonumber \\
\vdots \nonumber \\
W^{d-1}_1 & = & \sum_{i,\ldots,l,m=1}^d A^{d-2}_{1i}  \cdots A^0_{lm} W^0_m,
\label{simeq}
\end{eqnarray}
These are $d-1$ simultaneous equations for $W^0$.

The numerical procedure is as follows. We set the initial state of system
(\ref{SYS}) to a random initial guess vector $\left({\bf
y}^0\right)_{old}$ with $\left(y^0_1\right)_{old}=x^0_1$ and evolve it
using Eq.~(\ref{SYS}). Using this vector ${\bf y}(t)$ we write down $d-1$
simultaneous equations (Eqs.~(\ref{simeq})) which can be solved for $d-1$
unknown components of ${\bf w}^0=-\delta {\bf y}^0$.  Also, $\delta
y^0_1=0$. Thus the initial guess vector can be improved by
\begin{equation} 
\left( {\bf y}^0 \right)_{new} = \left( {\bf y}^0
\right)_{old} + \delta {\bf y}^0. 
\label{ITER} 
\end{equation} 
This sets up an iterative scheme giving us better and better estimates of
the initial vector which converge to ${\bf x}^0$.

We note that as in Newton-Raphson method, the choice of the initial guess
vector can be very important~\cite{PTVF}. In some cases, the iterative
procedure of Eq.~(\ref{ITER}) may not converge or converge to a wrong
root. In such cases, a different choice of initial guess vector can be
useful.

We further note the similarity of our method with the so called method of
variational equations in analytical dynamics~\cite{Whi}. The method of
variational equations can be applied to a known Hamiltonian system to
determine an unknown neighboring trajectory to an already known one.
There, the method requires a complete particular solution of a known set
of Hamiltonian equations of motion. In contrast, we have used our method
for dissipative chaotic systems. In such systems an analytical solution of
the equations of motion cannot be known. Further our method requires only
one component of a complete trajectory to be sampled. This is
important for the application of our method to secure communications
as we will demonstrate afterwards.

We now illustrate our method of estimating the initial state. As an
example we discuss the R\"ossler system given by~\cite{Ros}, $
{\bf \dot x} = (-x_2-x_3, x_1+ax_2, b+x_3(x_1-c))$.
First, we consider a case when the time series for $x_1$ is given and we
want to estimate $(x^0_2,x^0_3)$. We chose the parameters $(a,b,c)$ such
that the system is in the chaotic regime and the initial state ${\bf x}^0$
is on the chaotic attractor. We start with an arbitrary initial state
${\bf y}^0=(y^0_1,y^0_2,y^0_3)$ with $y^0_1 = x^0_1$. From
Eqs.~(\ref{simeq}) we get a pair of simultaneous equations for $(\delta y^0_2,\delta y^0_3)$,
which can be solved to obtain these values. With $\delta
y^0_1=0$ we use these in an iterative manner (Eq.~(\ref{ITER})) to obtain
the correct initial conditions. 

Let $e_i=|y^0_i-x^0_i|$ denote the absolute error in the
estimation of $x^0_i$. In Fig.~1(a) we plot a graph of 
errors $e_2$ and $e_3$ plotted
against the number of iterations, $n$, of our method~(Eq.~(\ref{ITER})). From
Fig.~1(a) we see that the errors go to zero and the successive estimates converge to the
correct values of $(x^0_2,x^0_3)$. Using only two data points in the given
time series $x_1(t)$, we can thus readily estimate the full initial state
${\bf x}^0$. We also note that the rate of convergence is very good. In
about 8 to 10 iterates we obtain the initial values $(x^0_2,x^0_3)$ to
within computer accuracy. If we write the deviations of the successive
iterates from the correct values in the form 
\begin{equation} 
(e_i)_n = \left| \left(y^0_i\right)_n - x^0_i \right| \sim e^{-\alpha n},
\label{converge} 
\end{equation} 
where $n$ is the number of iterations, then the value of the parameter
$\alpha$ is found to be $2.12$ for $e_2$ and $2.10$ for $e_3$. This is
consistent with the fact that Newton-Raphson method has a quadratic
convergence~\cite{Dev,PTVF}.

We note that the largest Lyapunov exponent for the subsystem
$(y^0_2,y^0_3)$ (conditional or subsystem Lyapunov exponent) is
positive~\cite{PC1}. The success of our method does not depend on whether
this Lyapunov exponent is positive or negative. This is important for
synchronization of chaotic signals.

We next present cases where time series for the variables $x_2$ and $x_3$
of the R\"ossler system are given. The procedure is similar to the case of
time series for $x_1$ as discussed above.  Fig.~1 (b) shows the errors
$e_1$ and $e_3$, when time series for $x_2$ is given, plotted against the
number of iterations. The parameter $\alpha$ (Eq.~(\ref{converge})) is
$2.08$ for $e_1$ and $2.10$ for $e_3$. This again indicates a quadratic
convergence. Similarly, Fig.~1 (c) shows the quantities $e_1$ and $e_2$
when time series for $x_3$ is given, as a function of the number of
iterations. The parameter $\alpha$ (Eq.~(\ref{converge})) is $2.02$ for
$e_1$ and 2.10 for $e_2$.
We note that the largest subsystem Lyapunov exponent is negative when time
series for $x_2$ is given and is positive when time series for $x_3$ is
given~\cite{PC1}. 

We have successfully applied our method to estimate the initial state
vector using a given scalar time series for many other dynamical systems
as well. These include Lorenz system~\cite{Lor} in its periodic, chaotic
or intermittent regimes, Chua's circuit~\cite{CKEI}, the disk
dynamo system modeling a periodic
reversal of earth's magnetic field~\cite{Rik,FW}, a 3-d plasma system
formed by a three wave resonant coupling equations~\cite{WFO} and a four
dimensional phase converter circuit~\cite{YK}.

We now turn to the important question of the rate of flow of
information about other variables into the given variable, i.e. the
variable whose time series is known, say $x_1$. To investigate this question we
start with the initial value $x_1(0)$ at $t=0$. This alone does not tell us
anything about the other variables. As time evolves information about
other variables flows into $x_1$ and the method presented above allows 
us to extract this information. As a specific example consider the
R\"ossler system with time series for $x_1$. Fig.~2 shows 
the plot of the asymptotic error $e_2$  of the estimation of $x_2(0)$
for large $n$, denoted by $e_\infty$, as a function of the total time
interval $\tau=(d-1) \Delta t$ for which the time series of $x_1$ is used in the
calculations. We observe that the accuracy of estimation improves with 
$\tau$ and shows a power law behaviour of the type
\begin{equation}
e_\infty = \tau^{-\nu}
\label{info}
\end{equation}
Fig.~2 shows a very good straight line behaviour on a log--log plot over seven
orders of magnitude and
the slope gives the value of the exponent $\nu = 2.00 \pm 0.01$.
The power law dependence of the accuracy on time interval as in 
Eq.~(\ref{info}) shows that the information entropy in terms of bits
or partitions grows logarithmically with time.
Note that this flow of information is different from the usual linear
growth of information entropy with time in chaotic systems which
corresponds to an exponential dependence of $e_\infty$ on $\tau$.
The excellent straight line fit over seven orders of magnitude that we
observe in Fig.~2, makes us believe that the power law dependence of
Eq.~(\ref{info}) and the corresponding logarithmic growth of
information content with time is a genuine property of dynamical systems. 
Further, the exponent $\nu$ appears to be universal
and has the same value within numerical errors
for the various examples that we have considered.

Now we will discuss an important application of our method for secure
communications. Let us suppose that Eq.~(\ref{SYS}) describes a
$d$-dimensional chaotic system at the transmitter end and a replica of 
the same at the receiver end. Let
$s(i), i=1,2,\ldots,k$ be the information signal to be encoded for 
communication and without loss of generality let $x_1(t)$ be the scalar
output signal to be transmitted. The signal $s(t)$ is encoded as
follows. While choosing the initial conditions of the vector ${\bf
x}(0)$, the component $x_1(0)$ is chosen arbitrarily while the other
$d-1$ components are chosen as $x_i(0) = s(i-1)$,
$i=2,\ldots,d$. Using the initial vector, ${\bf x}(0)$, we evolve
Eq.~(\ref{SYS}) for $d-1$ time steps of $dt$ to obtain the first $d$
values of the scalar variable $x_1$. This is the first stage of
evolution. The vector ${\bf x}(t_1)$, where $t_n=n(d-1)dt$, is
modified using $s(i)$ as explained below (Eq.~(\ref{MODIFY})) and the resultant is evolved
for $d-1$ time steps completing the second stage. The process is
repeated as often as required. After completion of each stage the
vector ${\bf x(t_n)}$ is modified as follows.
\begin{eqnarray}
x_1(t_n) & = & x_1(t_n) \nonumber \\
x_i(t_n) & = & x_i(t_n) + s(n(d-1)+i-1), \; \; i=2,\ldots,d.
\label{MODIFY}
\end{eqnarray}
Thus, assuming that $k$, the number of elements of information signal $s(i)$, is
divisible by $d-1$ we get $k+1$ values of the encoded signal $x_1$.

The recovery of the information signal at the receiver end using our
procedure of initial condition estimation is straightforward. At each
stage $d-1$ values of $x_1$ unable us to estimate the initial
conditions thereby recovering the signal from Eq.~(\ref{MODIFY}).

The communication procedure described above has several advantages.
Note that (a) The transmitted signal $x_1$ is not directly modulated by the
information signal $s$. (b) The information signal is added to more than 
one variables.
Due to these reasons, local approximation for the flow of the dynamical system in the
embedded phase space~\cite{ABST} cannot be used to attack the coding procedure.
 Secondly, the transmitted signal is of the
same size as that of the information signal, and there is no initial
transient as required in other synchronization based methods~\cite{CO}.
Thirdly,  as in other methods using chaotic encoding the same signal $s$ will give different
outputs for reruns of the coding procedure thereby making any attack
very difficult. 

Another important application of our initial condition estimation
method is in
the problem of synchronization of two identical chaotic systems coupled
unidirectionally by a scalar signal. 
The knowledge of initial conditions enables us to obtain immediate
synchronization thereby eliminating transients \cite{PC1}. Also,
the method could be used repeatedly to maintain synchronization. This
will be particularly useful for systems where the largest conditional
Lyapunov exponent is positive and other methods of synchronization fail~\cite{PC1}. 

To summarize, we have introduced a novel yet simple method to estimate
initial conditions of a multivariable dynamical system from a given scalar
signal. Our method is based on a multidimensional Newton-Raphson method
where we include the time evolution of the system. The method gives a
reasonably fast convergence to the correct initial state. The required
length of the time series is very small. The method works even when the
largest conditional Lyapunov exponent is positive. We find that the
information flow between different variables grows logarithmically
with time. This is different from the usual linear growth of
information in chaotic systems. An important application of the method 
is in secure communications. This method of communication has several
advantages as compared to other methods of communication based on
nonlinear dynamical systems.

\end{multicols}

\begin{figure}[tb]
\caption{ Plot (a) shows the errors $e_2$ and $e_3$ on a logarithmic
scale, as a function of $n$, the number of iterations of our method for
R\"ossler system when time series for variable $x_1$ is given. The pluses
show the values for $e_2$ and the crosses those for $e_3$. Errors are
seen to approach zero as $n$ increases.
The parameters are $(a,b,c)=(0.2,0.2,9.0)$, the
time step $\Delta t = 0.01$ and we use fourth order Runge-Kutta method
for the time evolution of R\"ossler equations. (We have also checked
our results with fifth and sixth order Runge-Kutta methods.) 
Similarly plot (b) shows the errors $e_1$ (pluses), and $e_3$
(crosses) as a function of $n$ when time series for $x_2$ is given.
Plot (c) shows the errors $e_|$ (pluses), and $e_2$ (crosses) as
a function of $n$ when time series for $x_3$ is given.}
      \label{fig1} 
\end{figure}

\begin{figure}[tb]
\caption{ The points in the graph show the asymptotic error $e_\infty$ of
initial state estimation as a function of $\Delta t$.
The case under consideration is the same as in
Fig.~1(a) for $e_2$ averaged over several realizations. A good straight line fit is obtained over seven
orders of magnitude and has a slope of $-2.0 \pm 0.01$. This indicates a logarithmic flow of
information from variable $x_2$  to variable $x_1$.}
      \label{fig2} 
\end{figure}

\psfig{figure=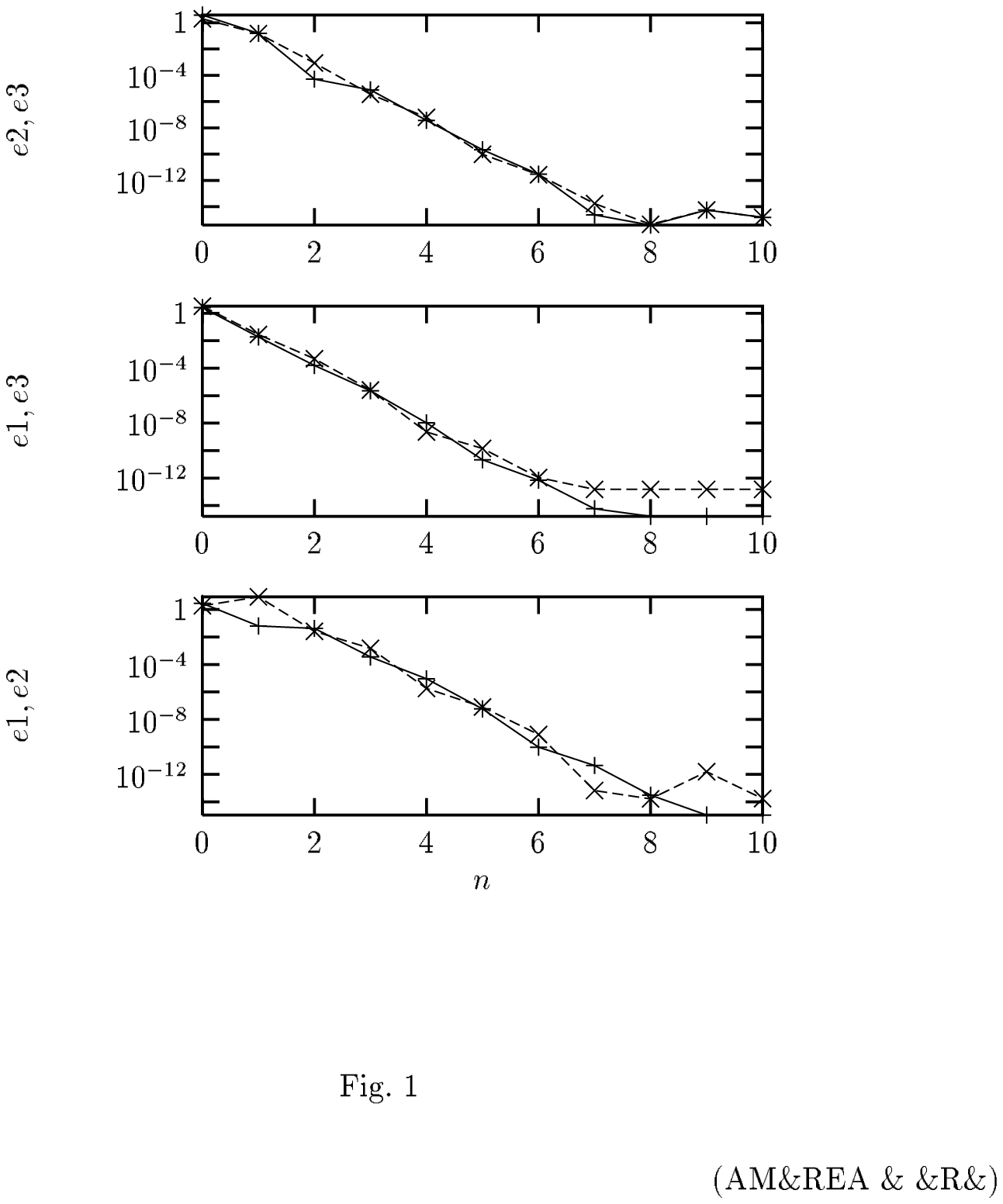}
\psfig{figure=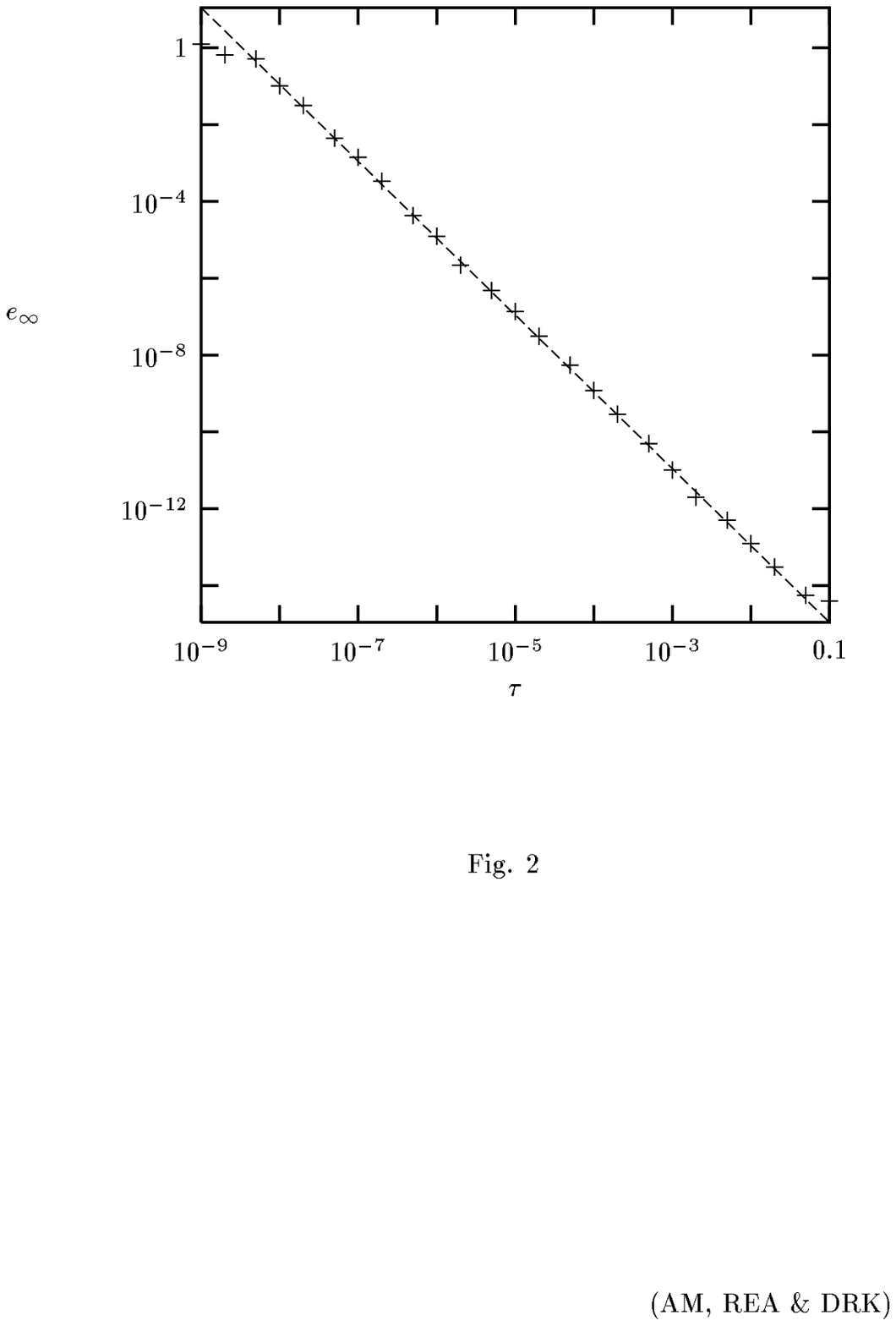}

\begin{references}

\bibitem{Dev} R.L. Devaney, {\it An Introduction to Chaotic Dynamical
Systems}, (The Benjamin / Cummings Pub. Co. Inc., Menlo Park - California,
1986).

\bibitem{PTVF} W.H. Press, S.A. Teukolsky, W.T. Vetherling and B.P.
Flannery, {\it Numerical Recipes in C}, 2nd ed. (Cambridge University, New
York, 1992), p. 379.

\bibitem{CO} K.M. Cuomo and A.V. Oppenheim, Phys. Rev. Lett. {\bf 71}, 65
(1993); K. John and R.E. Amritkar, Int. J. Bifurcation Chaos {\bf 4}, 1687 (1994);
U. Parlitz, L. Kocarev, T. Stojanovski, H. Preckel, Phys.
Rev. E {\bf 53}, 4351 (1996) and references therein.

\bibitem{HGO} S. Hayes, C. Grebogi and E. Ott, Phys. Rev. Lett. {\bf
70}, 3031 (1993).

\bibitem{PC1} L.M. Pecora and T.L. Carroll, Phys. Rev. Lett. {\bf 64}, 821
(1990).

\bibitem{Whi} E. T. Whittaker {\it A treatise on the analytical dynamics
of particles and rigid bodies}, $4^{th}$ Ed. (Dover publications, NY,
1944) p. 268.

\bibitem{Ros} O.E. R\"ossler, Phys. Lett. A {\bf 57}, 397 (1976).

\bibitem{Lor} E.N. Lorenz, J. Atmos. Sci. {\bf 20}, 130 (1963).

\bibitem{CKEI}L.O. Chua, L. Kocarev, K. Eckart and M. Itoh, Int. J.
Bifurc. and Chaos {\bf 2}, 705 (1992); also see Ref.~[6] p.238.

\bibitem{Rik} T. Rikitake, Proc. Cambridge Philos. Soc. {\bf 54}, 89
(1958).

\bibitem{FW} C. Flynn and N. Wilson, Am. J. Phys. {\bf 66}(8), 730 (1998).

\bibitem{WFO} J.M. Wersinger, J.M. Finn, E. Ott, Phys. Rev. Lett. {\bf
44}, 453 (1980).

\bibitem{YK} T. Yoshinaga and H. Kawakami in {\it Nonlinear dynamics in
circuits}, ed. L.M. Peccora and T.L. Carroll (World Scientific Pub. Co.
Pte. Ltd., Singapore, 1995) p. 111.

\bibitem{ABST} H.D.I. Abarbanel, R. Brown, J.J. Sidorowich, L.Sh.
Tsimring, Rev. of Mod. Phys. {\bf 65}, 1331 (1993).

\end{references}
\end{document}